\documentclass[a4paper,12pt]{article}

\usepackage[utf8]{inputenc}
\usepackage[english]{babel}
\usepackage[T1]{fontenc}

\usepackage{amsmath}
\usepackage{amssymb}
\usepackage{wasysym}
\usepackage{bm}

\usepackage{wrapfig}
\usepackage{color}

\usepackage{graphicx}
\usepackage[unicode]{hyperref}
\DeclareGraphicsExtensions{{.jpg},{.png},{.pdf}}
\usepackage{siunitx}
\usepackage{authblk}
\usepackage{xcolor}

\usepackage[margin=1in]{geometry}

\begin{document}
\sloppy

\title{Topological liquid crystal superstructures as structured light lasers}
\date{}
\author[1]{Miha Papi\v c\textsuperscript{\textdagger}}
\author[2]{Urban Mur\textsuperscript{\textdagger}}
\author[1]{Kottoli Poyil Zuhail}
\author[1,2]{Miha Ravnik}
\author[1,2]{Igor Mu\v sevi\v c}
\author[1,2,3]{Matja\v z Humar*}

\affil[1]{Department of Condensed Matter Physics, J. Stefan Institute, Jamova 39, SI-1000 Ljubljana, Slovenia}
\affil[2]{Faculty of Mathematics and Physics, University of Ljubljana, Jadranska 19, SI-1000 Ljubljana, Slovenia}
\affil[3]{CENN Nanocenter, Jamova 39, SI-1000 Ljubljana, Slovenia}
\affil[ ]{\textsuperscript{\textdagger}Equal contribution} 
\affil[ ]{*E-mail: matjaz.humar@ijs.si}

\maketitle

\textbf{
Liquid crystals (LCs) form an extremely rich range of self-assembled topological structures with artificially or naturally created topological defects. Some of the main applications of LCs are various optical and photonic devices, where compared to their solid state counterparts soft photonic systems are fundamentally different in terms of unique properties such as self-assembly, self-healing, large tunability, sensitivity to external stimuli and biocompatibility. Here we show that complex tunable microlasers emitting structured light can be generated from self-assembled topological LC superstructures containing topological defects inserted into a thin Fabry-Pérot microcavity. The topology and geometry of the LC superstructure determine the structuring of the emitted light by providing complex three dimensionally varying optical axis and order parameter singularities, also affecting the topology of the light polarization. The microlaser can be switched between modes by an electric field and its wavelength can be tuned with temperature. The proposed soft matter microlaser approach opens new direction in soft matter photonics research, where structured light with specifically tailored intensity and polarization fields could be designed and implemented.}\\

\vspace{0.5cm}

\noindent\fbox{%
	\parbox{\textwidth}{%
		\textbf{Significance}\\
		Liquid crystals (LCs) are used in a number of optical devices including lasers. However, till now only relatively simple LC structures have been employed inside laser cavities. LCs enable the self-assembly of extremely complex birefringent structures, which would be impossible to manufacture in any other way. LC molecules also orient the emission dipoles of the dye delivering efficient gain only to modes with the desired polarization. Further, due to sensitivity of the LCs to external stimuli they enable large tunability. This study provides experimental and simulation insights into coupling of light with the complex LC topological superstructures inside a laser cavity. This results in non-trivial intensity and polarization of the generated structured light.
	}%
}
\vspace{0.5cm}


The long range order and the fluid nature of LCs enable them to have some very unique properties not present in any other material. The large number of liquid crystalline phases in combination with inclusions and various boundary conditions results in a very wide variety of topological structures ~\cite{de1993physics}, which have been extensively studied in the last decades. Some of them form spontaneously, such as for example helices in cholesterics, 3D periodic structures in blue phases, focal conic domains in smectics, or are induced from outside, for example torons in frustrated cholesterics~\cite{smalyukh2010three}. By introduction of colloidal particles into the LCs~\cite{Musevicbook} or by confining LCs even more complex structures arise, including ordered colloidal crystals~\cite{smalyukh2005elasticity,muvsevivc2006two}, topological colloids~\cite{senyuk2013topological}, knotted structures~\cite{tkalec2011reconfigurable}, and complex director orientations in LC droplets and shells~\cite{lopez2011drops}. Active colloids~\cite{lavrentovich2016active} in LCs have also attracted great interest recently. Further, light-matter interaction can create for example optical solitons~\cite{assanto2012nematicons}.

Despite this richness of the LC structures only very few of them have been used inside a laser cavity till now. In most studies light is passed through a LC structure and in combination with other optical elements changes the light intensity, polarization and/or phase. A light field with custom designable intensity, polarization and phase fields is referred to as structured light~\cite{rubinsztein2016roadmap}. Structured light is highly relevant today, with applications ranging from imaging, metrology, optical trapping, ultracold atoms, to communications and memory~\cite{SWHell1994,parigi2015storage,ndagano2017creation,Lavery2017}. Special cases of structured light are optical vortex beams~\cite{shen2019optical,wang2018recent}, which carry orbital angular momentum and vector~\cite{Guzman2018,zhan2009cylindrical} beams with a spatially varying direction of polarization. 

Structured light is mostly created by passing a simple Gaussian beam through a phase plate, or q-plate, or by reflecting it from a metasurface, plasmonic nanostructure, or spatial light modulator ~\cite{rubinsztein2016roadmap,shen2019optical}.  For example, structured light can be generated by passing Gaussian light through a LC layer with a single topological defect of charge $q$~\cite{MarrucciPRL2006,rubano2019q}, through a nematic LC droplet with a $+1$ topological defect in its center~\cite{Brasselet2009}, through LC disclinations with charge ranging from $-3/2$ to $+3/2$~\cite{Loussert2013Brasselet} and even through more complex structures such as torons~\cite{yang2013arbitrary}. 

However, structured light can be also generated directly in the laser cavity~\cite{forbes2019structured,Naidoo2016}, which can be advantageous to achieve better mode purity and higher output powers. But a controllable and easy to manufacture microcavity that generates structured light on demand is still an open challenge. Examples include laser cavities with a spatial light modulator~\cite{ngcobo2013digital},  q-plate~\cite{Naidoo2016}, metasurface~\cite{metasurface} and microspheres~\cite{rivera2018fractal} inside the laser cavity and micro-ring cavities with broken mirror symmetry ~\cite{MiaoScience2016}. However, in most studies polarization of the laser output has been controlled in a limited way, primarily focusing on cylindrical vector vortex beams. There are also several examples of LC microlasers~\cite{coles2010liquid,zhang2020tunable,Yoshida2010,humar2009electrically,Jampani_fibers} which usually exhibit high tunability of the lasing wavelength, but not much emphasis has been given to the tailoring of the intensity profile of the output beam.

\begin{figure}
	\begin{center}
		\includegraphics[width=16cm]{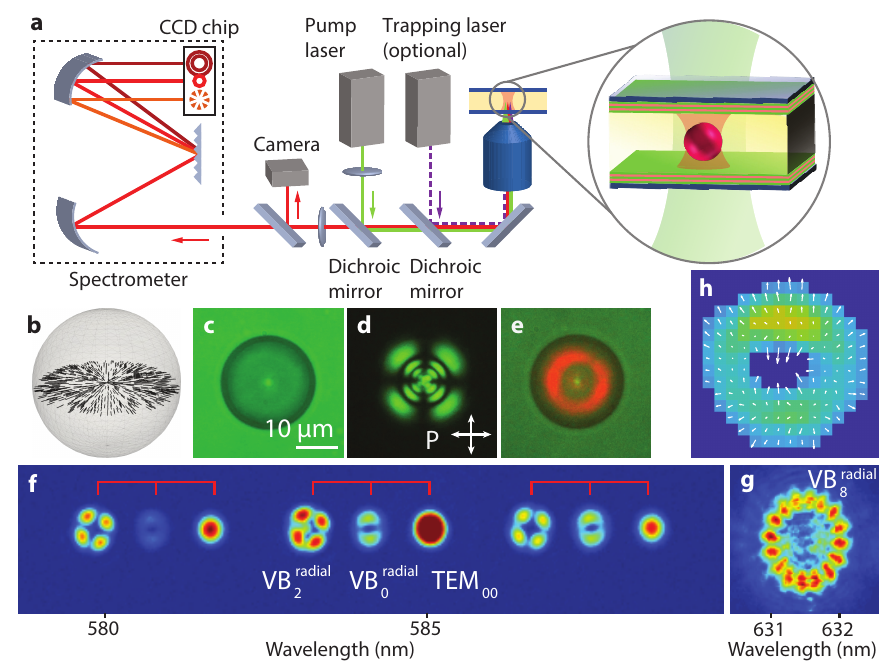}
		\caption{Scheme of optical setup and generation of laser vector beams. (a) Optical setup consists of a \SI{532}{nm} pulsed laser for optical pumping, an optional infrared \SI{1064}{nm} laser tweezers for manipulating the LC structures, an imaging spectrometer, a camera and an objective. The sample is composed of a LC structure, which is sandwiched in between two dichroic mirrors, forming a Fabry-Pérot microcavity. To analyze the optical modes, the LC laser output is sent through an imaging spectrometer with the slit wide-open. The image of the lasing emission pattern is spread in wavelength and imaged on a CCD sensor. The shape of each lasing mode is retained when passing through the spectrometer and the center position of each mode on the sensor indicates its wavelength. (b) Schematic representation of nematic director field (black) in the equatorial plane of a radial nematic droplet. (c) Bright-field image of a radial LC droplet in the cavity. (d) The same droplet viewed under crossed polarizers. (e) The droplet emits light when optically pumped above the lasing threshold. (f) Spectral decomposition of the laser modes emitted from another, smaller \SI{6.2}{\micro m} diameter radial nematic droplet in a \SI{25}{\micro m} cavity. Three transversal modes are supported:  VB$^{\mathrm{radial}}_2$, VB$^{\mathrm{radial}}_0$ and TEM$_{00}$ (simple Gaussian beam), which repeat as different longitudinal modes. (g) A single isolated mode, which is emitted from the droplet depicted in c-e with a diameter of \SI{21}{\micro m} in a \SI{25}{\micro m} cavity and is identified as VB$^{\mathrm{radial}}_8$. (h) A beam from a droplet analyzed with a wavefront sensor. The color of the pixels represents the intensity map and the arrows represent displacements of focal spots generated by the lenslet array.}
		\label{fig1}
	\end{center}
\end{figure}

In this work we demonstrate emission of laser vector beams with diverse and beyond standard intensity and polarization profiles generated from self-assembled LC superstructures with topological defects, confined into a Fabry-Pérot microcavity. The microcavity is made of two closely spaced dichroic mirrors. The cavity contains LC structures with a different topology, specifically the radial nematic microdroplets with a +1 point defect, nematic disinclination line defects, torons and cholesteric fingers in chiral nematic LC, colloidal chains and micro-tori, suspended in a nematic LC. The LC is doped with a fluorescent dye, which provides optical gain when illuminated with an external nanosecond pulsed laser. Above the lasing threshold, the microcavity emits coherent light, which is sent to a camera and an imaging spectrometer (Fig.~\ref{fig1}a). The topology of light is leadingly ascribed in the actual spatial profiles of the polarization of the emitted light modes. In the first part of the paper, laser modes from nematic droplets in an isotropic media are described, including mode polarization selection by dye orientation and mode switching and tuning by electric field and temperature. In the second part, more complex optical modes are shown as generated by chiral structures, specifically torons and chiral fingers. In the last part, vector beams with more general polarization are demonstrated by employing various LC structures with topological defects.

\subsection*{Results}
\textbf{Laser emission from LC droplets in a cavity.} We first demonstrate the emission of vector beams from a fluorescently doped radial nematic droplet  suspended in an isotropic medium filling the laser microcavity (Fig.~\ref{fig1}b-d). The droplet has a +1 topological point defect in the center, where the orientation of LC molecules is ill defined. A topological defect is characterized in 2D by a winding number (also called strength)~\cite{Musevicbook}, whereas in 3D it is characterized by a topological charge. The spherical shape of the droplet with a higher refractive index compared to its surroundings makes the Fabry-Pérot microcavity stable. The cavity configuration with a radial droplet is advantageous since the topological defect is always perfectly self-centered in the droplet and therefore also in the center of the generated laser beam. Above the lasing threshold of typically \SIrange{40}{100}{nJ} of pump pulse energy (Supplementary Section 1), laser light emission is observed as a red bright ring (Fig.~\ref{fig1}e).

In general, LC droplets can support also whispering gallery modes without the need of the external mirrors \cite{humar2011surfactant}. Whispering gallery modes are significantly different from the Fabry-Pérot modes, namely the light circulates near the surface of the droplet, which results in emission in all directions. Here the medium surrounding the droplets has a higher refractive index than in a typical case of whispering gallery mode lasers, which reduces the Q-factor of the droplet and suppresses the lasing of the whispering gallery modes, to achieve a more one-directional emission of the light. The Fabry-Pérot cavity filled with a gain material, but without any droplets, also emits laser light, but the output beam shape and polarization are not well determined. Also using the mirrors, forming the Fabry-Perot cavity is essential in the considered setup, since without the mirrors there would not be any well-defined modes or lasing, but only regular isotropic fluorescence emission.

By spectrally dispersing the generated laser light from a cavity containing a fluorescent droplet, we see that the emission is actually composed of several transverse laser optical modes, which repeat as different longitudinal orders (Fig.~\ref{fig1}f). The observed vector beams from laser cavities with embedded radial nematic droplet are identified as vector Bessel-Gauss-like (vBG) or vector Laguerre-Gauss-like modes (vLG), meaning that the beams are in principle unique, as determined by the actual liquid crystal structure embedded within the laser cavity, but can be approximately described with vLG or vBG profiles of zero total angular momentum (for more see Supplementary Section 2). The  observed modes are labeled as vector beams (VB) with radial, azimuthal, hybrid or linear polarization and mode number $n$ (VB$^\mathrm{radial/azimuthal/hybrid/linear}_n$), where $2n$ is the number of intensity maxima (or equally minima) in the azimuthal direction.

The mode number $n$ is dependent on the droplet size and cavity length with small droplets (below \SI{\sim 5}{\micro m}) emitting lowest order modes including simple Gaussian beams whereas larger droplets emitting petal-like modes with higher mode numbers (Fig.~\ref{fig1}g and Supplementary Section 3). The beams do not change with far field propagation (Supplementary Section 4) but are not always completely rotationally symmetric possibly due to a slight droplet deformation (Supplementary Section 5). It is important to stress that these beams are not vortex beams and do not carry orbital angular momentum, which can be clearly seen form the measurements of a single mode obtained by a Shack-Hartmann wavefront sensor (Fig.~\ref{fig1}h). The local projection of the Poynting vector to the plane perpendicular to the beam propagation does not have an azimuthal component (vorticity), therefore the beams do not carry orbital angular momentum \cite{leach2006direct}.

\begin{figure}[b!]
	\begin{center}
		\includegraphics[width=16cm]{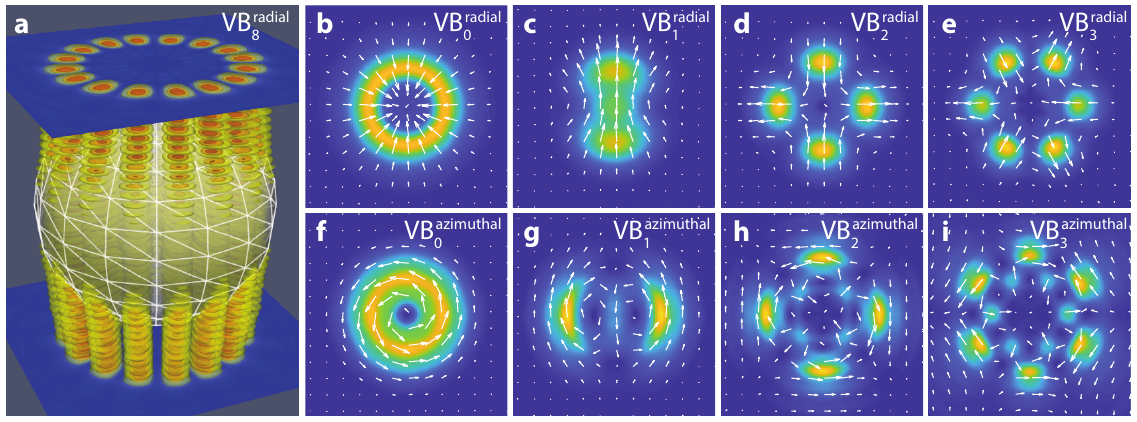}
		\caption{Numerically calculated resonant eigenmodes of a laser cavity containing a radial nematic droplet. (a) The droplet is placed in the center of the microcavity formed by two parallel, lossless and flat mirrors and assumed to be filled with an isotropic dielectric with a refractive index lower than the refractive indices of the LC. 3D representation of the intensity of VB$^{\mathrm{radial}}_8$ mode is plotted, where colors represent iso-surfaces of the normalized electric field intensity. (b-i) Polarization vectors (white arrows) and normalized intensity (color map) at the upper mirror plane. Four eigenmodes with the highest Q-factors are shown for (b-e) radial polarization, and (f-i) azimuthal polarization.}
		\label{fig2}
	\end{center}
\end{figure}

Finite-Difference Frequency-Domain (FDFD) approach was used to calculate standing lasing modes inside the microcavity containing dielectric media with spatially varying permittivity tensor (Supplementary Section 6). The calculated lasing profiles for highest Q-factor modes in a radial nematic droplet are presented in Fig.~\ref{fig2}, showing various modes which are similar in the longitudinal direction (Fig.~\ref{fig2}a), but with very distinct cross-section profiles (Fig.~\ref{fig2}b-i). The modes can be grouped into two families, depending on their general radial (Fig.~\ref{fig2}b-e) or azimuthal (Fig.~\ref{fig2}f-i) polarization. Each family consists of modes with different numbers of intensity maxima in the azimuthal direction, directly corresponding to the vector mode number. In an isotropic droplet all modes have a continuous intensity ring \cite{wu2019high} with the polarization axis direction continuously turning from radial to azimuthal. Here however, due to the birefringence of the LC, the polarization degeneracy is lifted and modes decouple into radial and azimuthal families at different frequencies. The modes with radial or azimuthal polarization are of effectively different size (with a cca. 10\% difference in the effective diameter of the beam), which is reflected in their relative Q-factors with effectively larger modes having smaller Q-factors, i.e.~higher losses.

\begin{figure}[b!]
	\begin{center}
		\includegraphics[width=11cm]{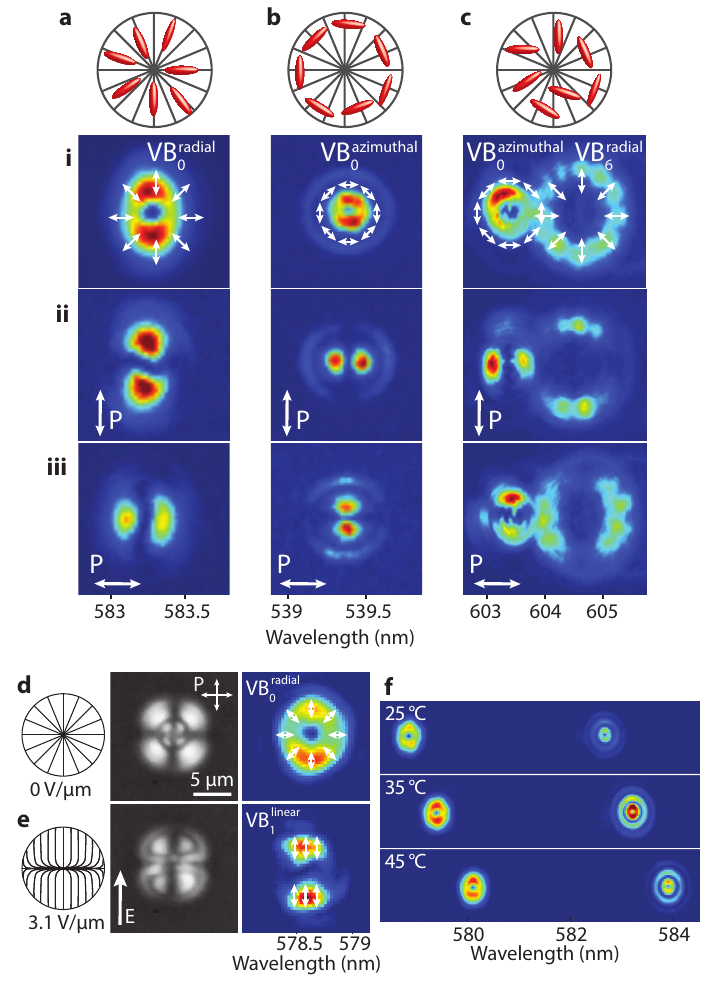}
		\caption{Experimentally demonstrated control of the vector beam polarization. (a-c) Vector modes generated from three radial nematic droplets, with three different orientations of the emission dipoles of the dye molecules: (a) along the nematic director; (b) perpendicular to the nematic director and (c) the emission dipoles are randomly oriented. (i) The intensity of vector modes in all three droplets, observed without polarizers; (ii) with a vertical polarizer and (iii) a horizontal polarizer. The white arrows denote the beam polarization directions and the P denotes the polarizer direction. (d) Director configuration (left), image under crossed polarizers (middle) and the generated radial vector beam (right) from a nematic droplet at zero electric field. (e) Same droplet with an applied electric field. (f) Two modes from radial nematic droplet at different temperatures.
		}
		\label{fig3}
	\end{center}
\end{figure}

\bigskip
\noindent\textbf{Polarization selection and mode switching.} In order to select the desired polarization in experiments, we need to provide more gain to that polarization. The use of LCs is advantageous, since the emission dipole of the dissolved dye can be oriented by the LC providing more efficient gain for that particular mode \cite{wright2004dye}. Depending on the particular dye the dipole can be aligned parallel or perpendicular to the optical axis (i.e.~the director) or can be unoriented (see Methods). In LCs very complex three-dimensional orientation and position arrangements of the gain dipoles can be achieved solely by self-assembly, realizing dipole emission which is strongly 3D anisotropic and nonhomogeneous, which would be very difficult to achieve in solid state lasers. 
With the dye oriented along the director (i.e.~radial direction), the radial component of the vector beam is generated with a larger amplitude than the azimuthal one (Fig.~\ref{fig3}a(i-iii)). Conversely, another dye with perpendicular orientation to the director (i.e.~azimuthal direction), generates an azimuthal polarization (Fig.~\ref{fig3}b(i-iii)). If the dye is randomly oriented, both polarizations are generated (Fig.~\ref{fig3}c(i-iii)). In the latter case typically the lower order modes will be azimuthally polarized, while higher orders will have a radial polarization, in line with the numerically calculated sizes (and relative Q factors) of the modes (Fig.~\ref{fig2}). It is important to note, that the birefringent nematic structures are embedded within the Fabry-Perot laser cavity, which effectively selects the main light propagation direction within the 3D nematic structure. For this reason, the light mainly experiences the effective 2D optical axis profile and symmetry within the transverse plane.

The major advantage of using LC optical materials is the ability to manipulate their optical axis  by external stimuli, which we use to control the output of the laser. The switching between modes can be achieved by applying an external in-plane electric field. Originally, a selected droplet emits radial vector beams (Fig.~\ref{fig3}d), but when an electric field is applied, the symmetry is broken and the optical axis in the droplet points mostly in the direction of the field (Fig.~\ref{fig3}e). The resulting mode is VB$_1^{\mathrm{linear}}$ (Supplementary Section 7). Additionally, to the electric field, the wavelength of modes can be tuned by changing the temperature, 
while notably, preserving the intensity and polarization profile of the beam (Fig.~\ref{fig3}f). The shift is caused by the change in the refractive indices and thermal expansion.

\begin{figure}[b!]
	\begin{center}
		\includegraphics[width=11.5cm]{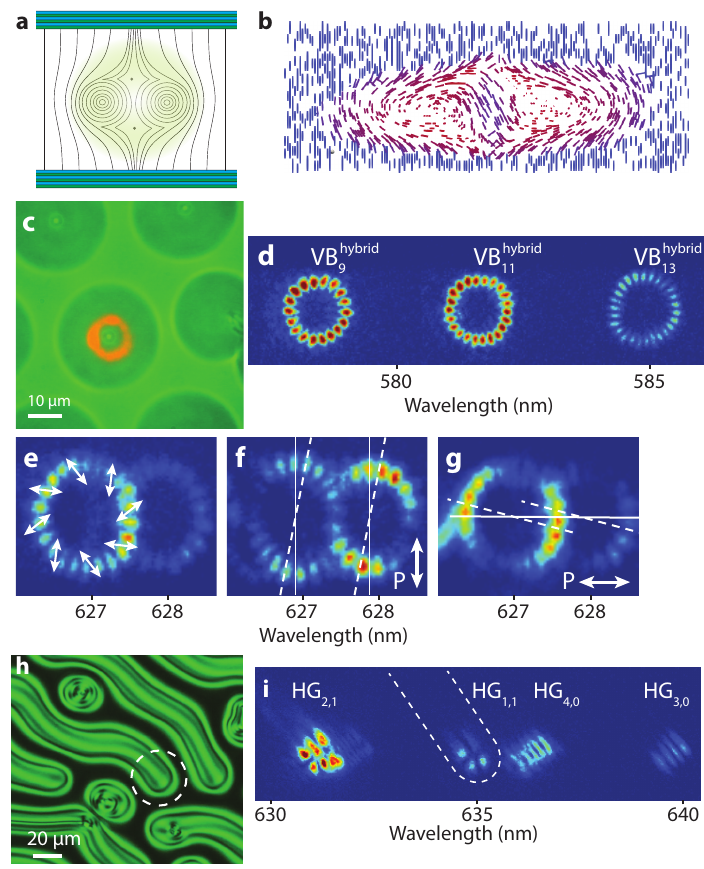}
		\caption{Modes emitted from chiral nematic structures. (a) Schematic representation of the director field in a toron. (b) The actual director field in the cross-section of a toron. Colors represent the orientation of nematic director with respect to the vertical direction. (c) A bright field image of several torons in a \SI{15}{\micro m}  microcavity. The central one is illuminated and pumped by an external laser and is emitting laser light, as seen from the bright red ring. (d) Spectral decomposition of the modes emitted from a toron in a microcavity. (e) Two selected modes emitted from a toron with the polarization direction indicated. (f) Same two modes with a vertical polarizer and, (g) a horizontal polarizer. The dashed lines are aligned with the areas of peak intensity. In all instances in Fig.~\ref{fig3} where pairs of polarizations are shown (vertical-horizontal), these pairs share the same intensity scale. All other panels have normalized intensities from the minimum to the maximum value. (h) Chiral fingers and torons under crossed polarizers. (i) A typical spectral decomposition of the laser output generated by a chiral finger, consisting of HG modes.
		}
		\label{fig4}
	\end{center}
\end{figure}

\bigskip
\noindent\textbf{Generation of modes from chiral nematic structures.} To further control the polarization of the modes, more complex self-assembled LC structures were employed. We first consider lasing from torons~\cite{Smalyukh2010} (Fig.~\ref{fig4}a,b). Torons are topologically protected LC structures, which are spontaneously formed in a layer of a cholesteric liquid crystal (CLC), with a thickness comparable to the helical pitch of the CLC. Torons have a triple-twisted structure, which is smoothly embedded in the homogeneous far-field director structure of a CLC. The droplet-like shape consists of a torus-like, triple-twisted chiral nematic structure, which has a topological charge of +2. For topological charge neutrality, two hyperbolic defects of charge -1 are present at the top and bottom of the torus, facing the flat interface, denoted by black dots. Because the thickness of the cavity is smaller than the pitch of the CLC, the CLC director profile is unwound almost everywhere except in places, where torons spontaneously appear. Similarly to larger nematic droplets, torons emit petal-like higher order vector modes (Fig.~\ref{fig4}c,d). The polarization of the modes is not completely radial but slightly angled at $\sim10^{\circ}$ (Fig.~\ref{fig4}e-g). These modes can be described as vector modes with a more general polarization direction, which is a linear combination of both radial and azimuthal polarization, therefore termed hybrid modes (Supplementary Section 7). This is a consequence of the complex triple-twisted structure of a toron, with the dipole moment of the fluorescent molecules following this twisted structure.

By selecting the shape of the surrounding high index region which focuses the light we can generate not only circular beam intensity profiles, but also other shapes. In order to generate Hermite-Gaussian (HG) modes cholesteric (chiral) fingers are used, which form when a toron is spontaneously elongated into a finger-like structure (Fig.~\ref{fig4}h). Due to their elongated shape these chiral fingers break the cylindrical symmetry of a toron and now emit HG beams (Fig.~\ref{fig4}i).

\begin{figure}[b!]
	\begin{center}
		\includegraphics[width=10.5cm]{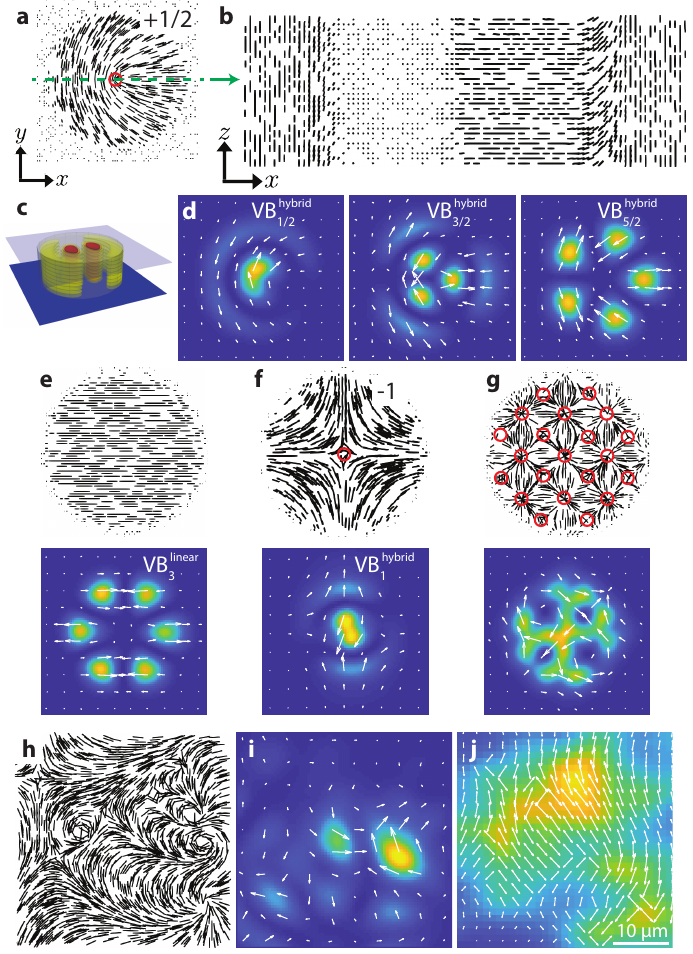}
		\caption{Simulated laser modes from LC topological defect line structures with different 2D topological charge (winding number). (a) A horizontal cross-section of LC defect line with a winding number of $+1/2$, spanning from one laser cavity reflective surface to another. This cross-section is the same at any position along the mode propagation. Defect line cores in all panels are indicated with red circles and the black lines indicate the optical axis (director). (b) A vertical cross section of the cylindrical region, containing a +1/2 defect line. (c) An example of the 3D intensity distribution of a representative mode (VB$^{\mathrm{linear}}_1$) generated by this kind of LC structures. (d) The $+1/2$ defect gives rise to hybrid VB modes with an odd number of maxima and the polarization following the direction of the nematic director. The first three modes with the highest Q-factors are shown. Calculated modes  (e) for a uniform optical axis (director) configuration, (f) for a defect with a winding number of -1 and (g) for an array of defects with  winding numbers of $\pm 1$. (h) A randomly oriented LC structure and (i) the simulated intensity and polarization maps of a lasing mode from this random structure. (j) Intensity and reconstructed polarization direction of experimentally realized lasing in a randomly oriented LC structure.}
		\label{fig5}
	\end{center}
\end{figure}

\bigskip
\noindent\textbf{Generation of complex vector beams.} To explore whether a complex director configuration can be designed to yield -in principle- a vector beam with an arbitrary polarization profile we have studied various other LC structures, focusing on different topologies of the nematic material. A nematic defect with a winding number of $+1/2$ was assumed to be in a cylindrical region, within the laser cavity (Fig.~\ref{fig5}a,b). The region acts as a waveguide and confines the modes with the polarization parallel to the director field (Fig.~\ref{fig5}c). The corresponding vector laser modes (for details see Supplementary Section 7) have intensity profiles with an odd number of minima and the polarization profile aligned with the direction of the underlying nematic director instead of a radial or azimuthal direction (Fig.~\ref{fig5}d), which shows an example how the topology of the nematic spatially varying optical axis can influence the topology of the laser light polarization (i.e.~the mode). This can also be clearly observed in the simple case of a homogeneous director orientation which yields a linear polarization (Fig.~\ref{fig5}e) but in a mode with non-trivial index $n$. Similarly, a $-1$ defect generates a hyperbolically shaped polarization profile (Fig.~\ref{fig5}f). Multiple $+1$ and $-1$ defects arranged into a square mesh have been also studied, creating multiple intensity zeros within the beam and the direction of the polarization twisting around these polarization singularities (Fig.~\ref{fig5}g).

Even further, both intensity and polarization can be generated in very complex profiles, regular and irregular. In order to show an example of a complex mode, a random distribution of $+1$ and $-1$ defects with a net topological charge 0 was constructed (Fig.~\ref{fig5}h). A laser mode with a complex intensity and polarization profile is generated in such a structure, which is conditioned by the spatially varying optical axis profile (Fig.~\ref{fig5}i). To corroborate the numerical simulations, we have experimentally fabricated a LC cell with a random anchoring direction on one surface so that also the LC is randomly oriented. The polarization of the beam generated by such a structure was reconstructed by capturing multiple polarization images (see Methods), revealing a complex spatial profile (Fig.~\ref{fig5}j). For liquid crystal structures, which emit multiple different transverse modes, a single mode can be excited by tightly focusing the pump beam and moving it across the structure. Such a selection of modes was demonstrated in colloidal structures assembled in liquid crystals (Supplementary Section 8) and chiral fingers (Supplementary Section 9).

\subsection*{Conclusions}
In conclusion, we have demonstrated that micro-lasers based on topological soft matter confined to a microcavity pave a new way towards compact and tunable microlaser sources of laser vector beams. We conjecture that it is in principle possible to create arbitrary laser beams by carefully designing appropriate 3D LC structures using reverse-engineering. The softness and fluidity of LCs as well as their adaptability to inserted objects enables the creation of complex 3D birefringent superstructures, which could be very difficult or even impossible to engineer by standard fabrication techniques such as solid-state lithography. And in turn, these complex three dimensional birefringent refractive index profiles allow for the creation of a large variety of modes with different polarization and intensity distributions. The self assembly properties of topological soft matter in combination with their large susceptibility to external fields might open new pathways to novel soft matter photonic devices. Finally, topological properties of LCs, which were explored in great depth during the last decade, could open the door to an entirely new class of microlasers, based on topological soft matter.

\subsection*{Materials and Methods}
\textbf{Sample preparation}. For the laser cavity planar dielectric mirrors with good transmission (>80\%) at the pump wavelength (\SI{532}{nm}) and high reflection ($\sim99.9\%$, OD3) in the range of maximum dye emission (\SIrange{550}{650}{nm}) were used. The mirror spacing was controlled by using sheet spacers, typically ranging from 5 to \SI{40}{\micro m}. In order to achieve planar or homeotropic anchoring, a 0.1\% solution of PVA (polyvinyl alcohol, Sigma) in water, or 1.5\% egg yolk lecithin (L-$\alpha$-phosphatidylcholine, Sigma) in diethyl ether were spin coated onto the surface, respectively. For non-degenerate planar anchoring the PVA coated substrates were also rubbed in a specific direction using a velvet cloth. In most experiments 5CB LC (4-cyano-4'-pentylbiphenyl, TCI), doped with 0.1\% Nile red (7-diethylamino-3,4-benzophenoxazine-2-one, Sigma) or 1\% Pyrromethene 580 (1,3,5,7,8-pentamethyl-2,6-di-n-butylpyrromethene-difluoroborate
complex, Exciton) was used. For random dye dipole orientation, 0.1\% Rhodamine B (Sigma) was used. For perpendicular orientation 0.1\% DiOC$_{18}$(3) (3,3'-Dioctadecyloxacarbocyanine perchlorate) in 8CB (4'-octyl-4-biphenylcarbonitrile) was used. In order to achieve sufficient levels of absorption of the infrared light from the laser tweezers, D77 dye ({5-[1,2,3,3a,4,8b-hexahydro-4-(4-methoxyphenyl)-cyclopeanta[b]indole-7-ylmethylene]-4-oxo-2-thioxo-thiazolidin-3-yl}acitic acid) in concentrations of up to 0.07\% was added. Cholesteric nematic LC mixtures were obtained by adding a right-handed chiral dopant CB15 ((S)-4-cyano-4'-(2-methylbutyl)biphenyl) ($\sim0.6\%$) to achieve the desired pitch ($p$). Different chiral structures were observed depending on the ratio $p/d$, where $d$ is the mirror spacing. Torons formed in homeotropic cells when $d/p \approx 0.7$ and chiral fingers were formed at slightly larger cell thickness. Radially oriented nematic LC droplets were created by suspending dye doped 5CB in a 1\% solution of egg lecithin in glycerol. Dispersions of colloids in the nematic LC were prepared from \SIrange{10}{15}{\micro m} BaTiO$_3$, \SI{10}{\micro m} borosilicate or \SI{5}{\micro m} silica colloids. Homeotropic anchoring on colloids was achieved by treating the particles with N,N-dimethyl-N-octadecyl-3-aminopropyl trimethoxysilyl chloride (DMOAP) silane. The torus-shaped colloids were produced by using  3D two-photon direct laser writing technique (Nanoscribe Photonic Professional) and treated for homeotropic anchoring as described elsewhere \cite{Uros_fract}. The tori had an outer diameter of \SI{50}{\micro m} and a thickness of \SI{5}{\micro m}. In plane electric field tuning was achieved by inserting two \SI{20}{\micro m} diameter parallel wires separated by \SI{200}{\micro m} into the cavity and an alternating voltage up to \SI{700}{V} peak to peak at \SI{1}{kHz} was applied.

\textbf{Optical setup}. In order to observe, excite and manipulate the samples, an optical setup built around an inverted microscope was used. A Q-switched doubled Nd:YAG laser with a wavelength of \SI{532}{nm}, pulse length of \SI{1}{ns}, maximum pulse energy of \SI{10}{\micro J}  (Pulselas-A-1064-500, Alpha-las) operated at repetition rates ranging from 3 to \SI{60}{Hz} was employed for excitation. A dichroic beamsplitter with high reflection below \SI{550}{nm} and high transmission above \SI{550}{nm} was inserted into the microscope filter turret. The laser was focused by a 60$\times$ 1.0 NA or 20$\times$ 0.50 NA objective. By placing a Galilean beam expander and additional lenses into the optical path before the dichroic beamsplitter, the laser beam was decollimated enabling us to vary the spot size on the sample from a sub-micron diffraction limited spot up to several hundred micrometers in diameter. The light emitted by the sample was collected by the same objective and sent to either a camera or a spectrometer. The imaging spectrometer (Andor Shamrock SR-500i), with a spectral resolution of up to \SI{0.05}{nm} was used. After the input had been separated by wavelength, it was recorded by a cooled back illuminated EM-CCD camera (Andor, Newton DU970N). The reconstruction of polarization was performed by taking intensity images with a polarizer inserted just before the camera at three different orientations: \SI{0}{\degree}, \SI{45}{\degree} and \SI{90}{\degree}. For every pixel of the image the polarization angle was calculated from the three measured intensities as \cite{wolff1995polarization}
\begin{equation}
	\theta=\frac{1}{2}\tan^{-1}\left( \frac{I_0+I_{90}-2I_{45}}{I_{90}-I_0}\right).
\end{equation}
Laser beams were characterized by a Shack-Hartmann wavefront sensor (Thorlabs, WFS40-7AR) as described before \cite{leach2006direct,stoklasa2014wavefront,grunwald2014spatio}. The wavefront sensor was placed at the output of the spectrometer. The shifts of the focal spots with a reference to a collimated beam were displayed as vectors. 
Wavefront tilt and defocus (corresponding to Zernike coefficients $Z_1^{-1}$, $Z_1^{1}$ and $Z_2^{0}$) were subtracted from the displacements. As a positive control a vortex beam was generated by sending a \SI{532}{nm} laser thought a spiral phase plate with a topological charge of one (RPC photonics, VPP-1b) and focused by a \SI{1000}{mm} focal length lens. Optionally, only for assembling the colloidal structures and moving defects in torons, the laser light from the optical tweezers (\SI{1064}{nm}) was focused onto the sample to a diffraction limited spot by using a second dichroic beamsplitter. The beam was steered with acousto-optic deflectors and controlled via computer.

\textbf{Numerical simulations} The Finite-Difference Frequency-Domain (FDFD) method was used in numerical simulations \cite{ivinskaya2011modeling}. Maxwell curl equations were written in the frequency domain and a matrix form of the equations was used. The eigenproblem for the magnetic field was formulated as:
\begin{equation}
\mu^{-1}\nabla \times \varepsilon^{-1} \nabla \times \bm{H} = \omega^2 \bm{H}
\end{equation}
The eigenvalue represents the frequency $\omega$ of the optical mode. The electric field $\bm{E}$ in every point of the cavity was calculated from the (nodal) eigenvector $\bm{H}$.  Information about the LC configuration between the mirrors of the FP microcavity was included in the electric permittivity matrix $\varepsilon$. Numerical results presented in this article were calculated by custom written code in the MATLAB R2019a environment and ran on Intel Xeon nodes with 190GB RAM. Perfectly matched layers (PML) were used to truncate the domain and simulate infinite boundary conditions in transverse directions. Perfect electric conductor boundary conditions were used on top and bottom boundaries to reproduce the effects of perfect mirrors. Quality factors were calculated directly from the real and imaginary part of $\omega$ and served as a tool to extract the localized modes and classify them. For more see Supplementary Section 6.

\subsection*{Acknowledgments}

This project has received funding from Slovenian Research Agency (ARRS) (P1-0099, N2-0085, N1-0104, J1-1697, N1-0116) and the European Research Council (ERC) under the European Union’s Horizon 2020 research and innovation programme (grant agreement No. 851143-Cell-Lasers and 884928-LOGOS).

\subsection*{Author Contributions}
M.H. designed the study. M.P. and K.P.Z. performed experiments and analyzed the experimental data. U.M. performed the numerical simulations. M.R., I.M. and M.H. supervised the project. All authors interpreted the results and prepared the manuscript.

\subsection*{Conflict of interest}
The authors declare no conflict of interest.

\bibliographystyle{unsrt}
\bibliography{references}

\end{document}